\documentclass[11pt]{article}
\usepackage{amsthm,amssymb, amsmath}
\usepackage[T1]{fontenc}

\usepackage{graphicx}
\newcommand{\InsertFig}[4]
{\begin{figure}[ht]
       \centerline{
         \includegraphics[width=#4]{#1}
       }
       \caption{{\footnotesize  #2}
       \label{fig:#3}}
\end{figure}}

\newcommand{\Tset}{\mathbb{T}}
\newcommand{\Zset}{\mathbb{Z}}

\newcommand{\Rset}{\mathbb{R}}

\newcommand{\cC}{\mathcal{C}}

\newcommand{\cF}{\mathcal{F}}

\newcommand{\cL}{\mathcal{L}}

\newcommand{\cR}{\mathcal{R}}
\newcommand{\cS}{\mathcal{S}}
\newcommand{\cT}{\mathcal{T}}

\theoremstyle{plain}
\newtheorem{teo}{Theorem}
\newtheorem{pro}[teo]{Proposition}
\newtheorem{lem}[teo]{Lemma}

\newtheorem{defi}{Definition}

\theoremstyle{remark}

\newenvironment{example}{\noindent \textbf{Example.}}{}

\newcommand{\pht}{\varphi_t}

\newcommand{\Sec}[1]{\S\ref{#1}}

\newcommand{\Pro}[1]{Prop.~\ref{#1}}

\newcommand{\Fig}[1]{Fig.~\ref{fig:#1}}
\newcommand{\Tbl}[1]{Tbl.~\ref{#1}}
\newcommand{\Def}[1]{Def.~\ref{#1}}
\newcommand{\Eq}[1] {(\ref{#1})}

\title{Generating Forms for Exact Volume-Preserving Maps}
\author{H.~E.~Lomel\'{\i} and J.~D.~Meiss \thanks
      {
        HL was supported in part by Asociaci\'{o}n Mexicana de Cultura, and
        JDM was supported in part by NSF grant DMS-0707659.
        Useful conversations with  Holger Dullin, Robert Easton
        and Rafael de la Llave are gratefully acknowledged. 
      }
    \\
 \begin{tabular}{cc}
	Department of Mathematics         &			Department of Applied Mathematics\\
    	Instituto Tecnol\'{o}gico Aut\'{o}nomo de M\'{e}xico  &University of Colorado \\
	Mexico, DF 01000 					&		Boulder, CO 80309-0526 \\
	lomeli@itam.mx  					&		James.Meiss@colorado.edu\\ 
\end{tabular}
	}
	
\begin{document}
\maketitle

\begin{abstract}
We study the group of volume-preserving diffeomorphisms on a manifold.
We develop a general theory of implicit generating forms. Our results generalize 
the classical formulas for generating functions of symplectic twist maps. 
\end{abstract}



\section{Introduction}

A map $f: M \to M$ preserves the volume form $\Omega$ in a manifold $M$  if
\[
	f^*\Omega = \Omega \;.
\]
For example, if $M = \Rset^n$ and the volume form is $\Omega = dx_1 \wedge dx_2 \wedge \ldots \wedge dx_n$, then $f$ is volume preserving when its Jacobian has unit determinant, $\det(Df) = 1$.

The study of such maps is interesting on one hand because volume-preserving maps are a simple and natural higher-dimensional generalization of the much-studied class of
area-preserving maps. On the other hand, the infinite dimensional group of volume-preserving diffeomorphisms on $\Rset^3$ is at the core of the ambitious program to reformulate
hydrodynamics~\cite{ArnoldK98}.
Volume-preserving maps arise in a number of applications such as the study of
the motion of Lagrangian tracers in incompressible fluids or of the structure
of magnetic field lines~\cite{Holmes84, LauF92, SotiropoulosVL01,MullowneyJM05}.

In this paper we will study the construction of \emph{generating forms} for exact volume-preserving maps. A similar construct, \emph{generating functions}, is familiar in the exact symplectic case. Recall that a symplectic map $f$ preserves a nondegenerate, closed two-form $\omega = dq \wedge dp$ defined on an $n=2d$ dimensional manifold: $f^*\omega = \omega$.
Exact symplectic maps arise when $\omega$ is exact. For instance when there exists a \emph{Liouville one-form} $\nu$ on a cotangent bundle, the symplectic form is defined by $\omega=-d\nu$. Then $f$ is an exact symplectic map if
\begin{equation}\label{eq:symplecticGen}
	f^*\nu-\nu=dL \;,
\end{equation}
for a ``generating'' function $L$ defined on $M$. If we denote
the map by 
\[
	(Q,P) = f(q,p) \;,
\]
and choose, e.g.,  $\nu = p dq \equiv \sum_{i=1}^{d} p_i dq_i$, then \Eq{eq:symplecticGen} has the form
\begin{equation}\label{eq:symplecticII}
	PdQ - pdq = dL \;.
\end{equation}
The theory of canonical generating functions regards this as an equation not on $M$, but on the doubled phase space $N = M \times M$ with coordinates $(q,p,Q,P)$ \cite{Abraham78,Easton98}. In this case, \Eq{eq:symplecticII} is not valid everywhere on $N$, but only on the graph $F = \{(q,p,Q,P) \;|\; (Q,P) = f(q,p)\} \in N$ of $f$. Alternatively, we say that $L:N \to \Rset$ is a generating function, with respect to $\nu$, if the set  on which the one-form \Eq{eq:symplecticGen} vanishes is precisely the graph $F$. The resulting map $f$ is necessarily exact symplectic.

There are four special cases of \Eq{eq:symplecticGen} that are typically defined \cite{Goldstein02}. For example, if $L$ is assumed to depend only upon $(q,Q)$, then \Eq{eq:symplecticII} is equivalent to the implicit equations
\begin{align*}
	p &= -\partial_q L(q,Q) \;,\\
	P &=  \partial_Q L(q,Q) \;.
\end{align*}
These generate a map when the implicit equations can be solved for $(Q,P)$; this occurs under a \emph{twist condition}, $\partial_{qQ} L(q,Q) \neq 0$, that is, the matrix of partial derivatives $\partial_{q_iQ_j} L(q,Q)$ is either positive or negative definite. Many other generating functions can be obtained by other choices of the form $\nu$ \cite{Arnold78}. 

In the following sections, we analyze the group of exact volume-preserving diffeomorphisms and obtain implicit generating \emph{forms} for exact volume-preserving maps by mimicking the symplectic construction. In particular, in some cases it is possible to determine an exact volume-preserving diffeomorphism $f$ from an $(n-2)$-form $\Lambda$ on $N = M \times M$.

We start with a discussion of exact volume-preserving maps in \Sec{sec:exact}. Examples of generating forms were first given---as far as we know---by Carroll \cite{Carroll04}, see \Sec{sec:Carroll}, though he did not use the notation of differential forms. The general formulation is given in \Sec{sec:generating}, and additional examples are presented in the following sections. A volume form is not always exact, for example when $M = \Tset^n$. However, in some cases a generating form can still be obtained on the universal cover of $M$ as we discuss in \Sec{sec:manifolds}. Applications to maps on $\Tset^d \times \Rset^k$ are given in the last section.

\section{Exact Volume-preserving maps} \label{sec:exact}

A volume form $\Omega$ is exact when there exists an $(n-1)$-form $\alpha$ such that $\Omega= d\alpha$. For this case, exact volume-preserving maps can be defined by analogy with the symplectic case.

\begin{defi}\label{def:exact}
Let $(M,\Omega)$ be a manifold in which the volume form $\Omega$  is exact. Suppose that $d\alpha = \Omega$.
A diffeomorphism $f:M\to M$ is \emph{exact} volume preserving if there exists  an $(n-2)$-form $\lambda$ on $M$ such that
\begin{equation}\label{eq:exact}
 f^*\alpha-\alpha = d\lambda.
\end{equation}
We will denote by $\mathrm{Diff}_{\alpha}\,(M)$ the set of exact volume-preserving diffeomorphisms with respect to $\alpha$.
\end{defi}

\noindent
It is clear that if $f$ is exact volume preserving, then $f^{-1}$ is also.  Moreover, if $f = f_1 \circ f_2$ is the composition of exact volume-preserving maps with forms $\lambda_1$ and $\lambda_2$, respectively, then since $(f_1 \circ f_2)^* = f_2^*  f_1^*$,
\[
	f^*\alpha-\alpha   = f_2^* (f_1^*\alpha -\alpha) + f_2^*\alpha -\alpha 
	   = d(f_2^*\lambda_1 +\lambda_2) \;.
\]
Thus $f$ is exact volume-preserving with
\begin{equation}\label{eq:composition}
	\lambda = f_2^*\lambda_1 + \lambda_2 \;.
\end{equation}
Therefore $\mathrm{Diff}_{\alpha}\,(M)$ is a group that can be regarded as an infinite dimensional Lie group. Clearly, if we have two forms $\alpha$ and $\tilde{\alpha}$, for which $\alpha-\tilde{\alpha}$ is exact, then 
$\mathrm{Diff}_{\alpha}\,(M)=\mathrm{Diff}_{\tilde{\alpha}}\,(M)$.


When $f$ is exact volume preserving, the form $\lambda$ can be used to compute volumes of invariant or partially invariant regions. For example, suppose that $\cC$ is an orientable, boundary-free, codimension-two manifold that is invariant under $f$, e.g, if $\dim{M} = 3$ then $\cC$ is an invariant circle. Let $\cS$ be any codimension-one embedded submanifold bounded by $\cC$ and $\cR$ be the ``region'' bounded by $\cS$ and its image, $\partial \cR = f(\cS) - \cS$. In other words, suppose that a region $\cR$ is bounded by $\cS$ and its image, with the appropriate orientations. The (algebraic) volume of $\cR$ is 
\[
	Vol(\cR) = \int_\cR \Omega = \int_\cS f^*\alpha-\alpha   = \int_{\cC} \lambda \;.
\]
Generalizations of this formula can also be used to compute the flux of orbits escaping from a resonance zone in terms of the integral of the form $\lambda$ along heteroclinic intersections of stable and unstable manifolds \cite{Lomeli08b}. Similar formulas have been extensively used in the symplectic case \cite{MMP84, MMP87, Easton91} and should prove useful in studies of volume-preserving transport \cite{Piro88, Feingold89, Balasuriya05}.

Any exact symplectic map of a two-dimensional manifold is exact volume-preserving with the volume form $\Omega = \omega$, if we choose the one-form $\alpha=-\nu$ and the zero-form $\lambda= -L$. This also holds more generally.

\begin{lem}
Any exact symplectic diffeomorphism is exact volume-preserving.
\end{lem}

\proof
Let $f:M\to M$ be an exact symplectic diffeomorphism. The $d$-fold wedge of the two-form $\omega$ is a volume form\footnote
{
	The standard volume would be $\frac{(-1)^{\lfloor{d/2}\rfloor}}{d!} \Omega$.
}
\[
	\Omega=\omega^{\wedge d} \equiv \underbrace{\omega\wedge \omega\wedge \cdots\wedge \omega}_{d}.
\]
Defining $\alpha = -\nu \wedge \omega^{\wedge d-1}$ then $d\alpha=\Omega$, and
\begin{align*}
  f^*\alpha-\alpha &= f^*\left(-\nu\wedge \omega^{\wedge d-1} \right)+ \nu\wedge \omega^{\wedge d-1}
	                = -(f^*\nu-\nu)\wedge \omega^{\wedge d-1}\\
	               &= - dL\wedge \omega^{\wedge d-1}= d(-L \omega^{\wedge d-1}) \;,
\end{align*}
since $f^*\omega = \omega$ and $d\omega = 0$. Thus $f$ is exact volume preserving with the $2(d-1)$ form $\lambda = -L\omega^{\wedge d-1}$.
\qed

\subsection{Exact Incompressible Vector Fields}

One important aspect of the structure of Lie groups is the study of their one-parameter subgroups. Here we consider the subgroups of $\mathrm{Diff}_{\alpha}\,(M)$ generated by exact incompressible vector fields. Recall that an incompressible vector field  $X$ satisfies $L_X\Omega \equiv (\nabla \cdot X) \Omega = 0$, where $L_X$ is the Lie derivative. In other words, $X$ is incompressible if and only if the corresponding flow $\pht$ generated by $X$ is volume-preserving for each $t$. To find a similar condition for exact volume-preserving flows, suppose that the flow $\pht$ is an exact volume-preserving diffeomorphism for each $t$. According to \Def{def:exact} there exists a smooth family of $(n-2)$-forms $\lambda_t$ such that
\begin{equation}\label{exactFlow}
    \pht^*\alpha-\alpha= d\lambda_t \;.
\end{equation}
Differentiating with respect to $t$ gives
\begin{align*}
      \frac{d}{dt}\pht^*\alpha
      		&= \pht^*L_X\alpha\\
			&= \pht^*\left(i_Xd\alpha+di_X\alpha\right)
			 = d\left(\frac{\partial}{\partial t}\lambda_t\right) \;.
\end{align*}
Since $d\alpha = \Omega$, this gives $i_X\Omega= d\left(\varphi_{-t}^*\frac{\partial}{\partial t}\lambda_t -i_X\alpha\right)$; consequently, the flow generated by the vector field $X$ is exact volume-preserving if and only if $i_X\Omega$ is exact. 

We will argue that the expression $d\left(\varphi_{-t}^*\frac{\partial}{\partial t}\lambda_t\right)$ does not depend on time. Using the group property of the flow $\pht$  in \Eq{exactFlow}, it follows that for all $t,s \in \Rset$.
\begin{equation}\label{eq:chida}
	d\lambda_{s+t}=d\left(\varphi_s^*\lambda_{t}+ \lambda_{s} \right) \;.
\end{equation}
Letting $\beta=\left.\frac{\partial}{\partial t}\right|
_{t=0}\lambda_t$, then differentiating \Eq{eq:chida} with respect to $t$ and setting
$t = 0$ gives
\begin{equation}\label{lambdaS}
	d\left(\frac{\partial\lambda_s}{\partial s}\right)
	= d\left(\varphi_s^*\beta\right).
\end{equation}
Therefore $d\left(\varphi_{-t}^*\frac{\partial}{\partial t}\lambda_t\right)=d\beta$ and  $L_X\alpha= d\beta$. Consequently $i_X\Omega= d\left(\beta - i_X\alpha\right)$ is exact. 
Moreover, \Eq{lambdaS} shows that the form
\[
 \lambda_t-\int_0^t \varphi_\tau^*\beta\,d\tau
\]
is closed and, without loss of generality, we can choose
\begin{equation}\label{lambdaT}
 \lambda_t=\int_0^t \varphi_\tau^*\beta\,d\tau \;.
\end{equation}
We summarize these results as a proposition.

\begin{pro}\label{prop:liouville}
 Let $X$ be a vector field on a smooth manifold $M$ of dimension $n$
 with an exact volume form $\Omega$, such
that $\Omega=d\alpha$, for some fixed $(n-1)$-form $\alpha$.
 Let $\pht$ be the flow generated by $X$, and suppose that it is complete.
Then the following are equivalent.
\begin{enumerate}
	\item $i_X\Omega$ is exact.
	\item There exists an $(n-2)$-form $\beta_X$ such that $L_X\alpha= d\beta_X$. 
	\item For each $t\in \Rset$, there exists an $(n-2)$-form $\lambda_t$, \Eq{lambdaT}, such 
	that \Eq{exactFlow} is satisfied.
\end{enumerate}
\end{pro}

If a vector field on a smooth manifold $M$ satisfies any of the conditions of \Pro{prop:liouville}, we will say that $X$ is an \emph{exact incompressible} vector field with structure form $\beta_X$. Another term that has been used for these vector fields is ``globally Liouville,'' see for example \cite{Gaeta03}. These vector fields have a Lie algebraic structure.

\begin{lem} If $X,Y,Z$ are exact incompressible vector fields, with structure forms $\beta_X, \beta_Y$ and $\beta_Z$ then,
\begin{enumerate}
\item the Lie bracket $[X,Y]$ is an exact incompressible  vector field with structure form 
\begin{equation}\label{eq:structure1}
	\beta_{[X,Y]}=L_X\beta_Y-L_Y\beta_X
\end{equation}
and,
\item if $f$ is exact volume-preserving with $f^*\alpha-\alpha = d\lambda$, the pull-back $f^*Z$ is an exact incompressible  vector field  with structure form 
\begin{equation}\label{eq:structure2}
	\beta_{f^*Z}=f^*\beta_Z-L_{f^*Z}\lambda \;.
\end{equation}

\end{enumerate}

\end{lem}

An interesting exercise is to check that the formulas (\ref{eq:structure1}) and (\ref{eq:structure2}) are compatible. Since
$f^*[X,Y]=[f^*X,f^*Y]$, it must be the case the form $\beta_{f^*[X,Y]}-\beta_{[f^*X,f^*Y]}$ is closed. In fact it is possible to show that $\beta_{f^*[X,Y]}=\beta_{[f^*X,f^*Y]}$ directly from the lemma. 

\begin{example}
Consider the case of a nonautonomous Hamiltonian flow, generated by a $C^2$ function $H:\Rset^3\to\Rset$. With $H(q,p,\theta)$ we form the autonomous Hamiltonian vector field given by
\begin{equation}
    X_H = ( H_p, -H_q,1)^T \;.
\end{equation}
The volume form is $\Omega= dq\wedge dp \wedge d\theta$ and we can choose
$\alpha= -p dq \wedge d\theta $ so that $d\alpha=\Omega$. Hence,
\begin{align*}
	i_{X_H} \Omega &= dH \wedge d\theta + dq \wedge dp \;, \\
	i_{X_H} \alpha &= -p\frac{\partial H}{\partial p}d\theta + p dq \;.
\end{align*}
These imply
\[
 	L_{X_H}\alpha=i_{X_H}\Omega + di_{X_H}\alpha 
		     =d\left(H-p\frac{\partial H}{\partial p} \right)\wedge d\theta \;. 
\]
so that we can define $\beta= - \cL_H d\theta $ where 
\begin{equation}\label{lagrangian}
	\cL_H=pH_p-H
\end{equation}
is the Lagrangian, and
\begin{equation}
	\lambda_t= - \left(\int_0^t \cL_H\circ\varphi_\tau\,d\tau\right) d\theta \;.
\end{equation}
It is possible to show directly from equation (\ref{eq:structure1}) that
\[
\beta_{[X_H,X_G]}=- \cL_{\{G,H\}} d\theta, 
\]
where $\{G,H\}=G_qH_p-G_pH_q$ is the Poisson bracket.

\end{example}

\section{Generating Forms: Carroll's Example}\label{sec:Carroll}

In this section, we consider the simplest case where $M = \Rset^3$ and $\Omega =dx \wedge dy \wedge dz$. The map $(X,Y,Z) =f(x,y,z)$ is volume-preserving when $f^*\Omega = dX\wedge dY \wedge dZ = \Omega$. Here we give a simple example, based on that of Carroll \cite{Carroll04}, of an implicit generating form for $f$ when it is exact.

The map $f$ is exact volume-preserving when it satisfies \Eq{eq:exact}; however, this notion can be slightly generalized by noting that there are many choices for the form $\alpha$ such that $\Omega = d\alpha$, and we can use different representatives for the two forms in \Eq{eq:exact}, $f^*\alpha$ and $\alpha$. Indeed, letting $\tilde\alpha$ and $\alpha$ be two different such representatives, we can generalize \Eq{eq:exact} to
\begin{equation}\label{eq:generating}
	f^*\tilde\alpha -\alpha = d\lambda \;.
\end{equation}
Since $\tilde\alpha - \alpha$ is closed and every closed form on $\Rset^3$ is exact, this is equivalent to \Eq{eq:exact} (however, for more general manifolds some additional care must be taken, see \Sec{sec:manifolds}).

There are three simple natural choices for $\alpha$: $x dy\wedge dz$, $y dz \wedge dx$, and $z dx \wedge dy$. Therefore for $\alpha$ and $\tilde\alpha$, there are together, nine possible choices. Essentially we are choosing a subset of the variables $(x,y,z, X,Y,Z)$---some ``old'' and some ``new''---and combining them in a single function.
We will select the one-form $\lambda$ to depend explicitly on the  variables chosen.

For example let us choose $\tilde\alpha=zdx\wedge dy$ and $\alpha=x dy\wedge dz$ and try to find a diffeomorphism $f(x,y,z)=(X,Y,Z)$ such that
\[
	f^*\tilde\alpha -\alpha = Z dX \wedge dY -x dy\wedge dz=d\lambda \;, 
\] where $\lambda$ is a one-form, say,
\begin{equation}\label{eq:carroll}
	\lambda = \Phi(y,z,X,Y)dy + \Psi(y,z,X,Y)dY \;.
\end{equation}
Here $\lambda$ is to be thought of as a one-form on $\Rset^3$; that is, it must be evaluated on the transformation $(X,Y,Z) = f(x,y,z)$. However, we ignore that for the moment and treat $(y,z,X,Y)$  as four independent variables. The differential of \Eq{eq:carroll} is
\[
d\lambda = \partial_z\Phi dz \wedge dy + \partial_X\Phi dX \wedge dy +
 							(\partial_Y\Phi-\partial_y\Psi)dY\wedge dy + 
		\partial_z\Psi dz \wedge dY + \partial_X\Psi dX \wedge dY \;.
\]
To be consistent with \Eq{eq:generating} $\Phi$ must independent of $X$, $\Psi$ independent of $z$, and
\begin{equation}\label{eq:carrollMap}
	x = \partial_z\Phi(y,z,Y)\;, \quad 
	\partial_Y\Phi(y,z,Y) = \partial_y\Psi(y,X,Y) \;, \quad
	Z = \partial_X\Psi(y,X,Y) \;.
\end{equation}
These three equations locally define a map $f$, provided that the first equation can be inverted for $Y$, 
which requires that $\partial_{zY}\Phi \neq 0$, and that the second can be solved for $X$, which requires $\partial_{yX}\Psi \neq 0$. By analogy with the case of symplectic maps, we call these conditions \emph{twist conditions}.
Note that the twist conditions are geometrical properties of $f$ and $f^{-1}$, namely
\begin{equation}\label{eq:carrollTwist}
	\frac{\partial Y}{\partial x} \neq 0  \;, \quad 
	\frac{\partial y}{\partial Z} \neq 0 \;.
\end{equation}
The map $f$ is globally defined if for each $(y,z)$ the image of the line $L_{y,z} = \{(s,y,z) \;|\; s\in \Rset\}$ intersects every plane $P_Y = \{(u,Y,v) \;|\; (u,v) \in \Rset^2\}$ exactly once. Conversely, for each $(X,Y)$ the preimage of the line $L_{X,Y} = \{(X,Y,s) \;|\; s \in \Rset\}$ intersects every plane $P_y$ exactly once.  This is shown in \Fig{twistConditions}. 
Consequently, not every exact volume-preserving map has a generating form of this type.

\InsertFig{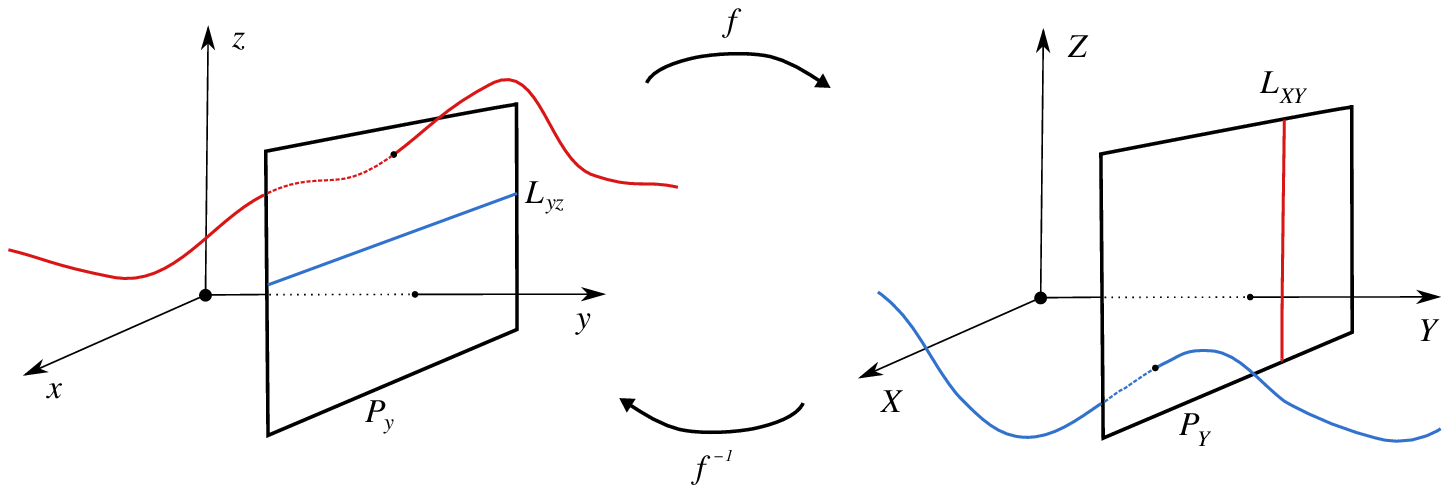}{Illustration of the twist conditions for the map \Eq{eq:carrollMap}}{twistConditions}{6in}

As an example, let $\Phi(y,z,Y) = zY+ g(y,Y)$ and $\Psi(y,X,Y) = Xy $. Then the map generated by \Eq{eq:carrollMap} is
\[
	(X,Y,Z) = (z + \partial_2 g(y,x) ,x,y) \;,
\]
which is of the form of the shift-like diffeomorphisms studied in \cite{Lomeli98a, Bedford98a, Li04, Gonchenko06}. It trivially satisfies the twist conditions since $Y(x,y,z) = x$ and $y(X,Y,Z) = Z$.

\section{Generating Forms}\label{sec:generating}

Though we thought of $\lambda$ in \Sec{sec:Carroll} as a form on $M = \Rset^3$, it is more properly thought of as a form on the product space $N = M \times M$. Thus if $(x,y,z,X,Y,Z)$ are the coordinates of a point in $N$, the expression \Eq{eq:carroll} becomes a one form on $N$. To distinguish this new form from the original form on $M$, we will call it $\Lambda$.

More generally, let $\tilde \alpha$ and $\alpha$ be two $(n-1)$-forms such that $d\alpha=d\tilde{\alpha}=d\Omega$. Let $\pi_{1,2}:N\to M$ be the projections
\begin{equation}\label{eq:project}
	\pi_1(m_1,m_2)=m_1 \;, \quad \pi_2(m_1,m_2)=m_2 \;.
\end{equation}
Following the symplectic case \cite{Abraham78, Easton98}, a generating form will be constructed using $\pi_2^*\tilde{\alpha}-\pi_1^*\alpha$, which is an $(n-1)$-form on $N$. Note that if $(\xi_1,\xi_2) \in T_{(m_1,m_2)}N$, then $(\pi_2^*\tilde{\alpha}-\pi_1^*\alpha)_{(m_1,m_2)}(\xi,\eta) = \tilde\alpha_{m_2}(\xi_2) - \alpha_{m_1}(\xi_1)$. 

\begin{defi}[Generating Form]\label{def:generating}
An $(n-2)$-form $\Lambda$ on $N=M\times M$ is a \emph{generating form} with respect to the pair $(\alpha,\tilde{\alpha})$ if the set $F \subset N$ on which the form
\begin{equation}\label{eq:Gamma}
	\Gamma \equiv \pi_2^*\tilde{\alpha}-\pi_1^*\alpha - d\Lambda
\end{equation}
vanishes is the graph $F = \{ (m,f(m)) \;|\; m\in M \}$ of a smooth function $f:M\to M$. In this case, we will say that the map $f$ is \emph{generated} by $\Lambda$.
\end{defi}

The idea of using a form to define a submanifold is very old. It goes back to the question of solving Pfaffian equations to define subbundles of a vector bundle, in particular of the tangent bundle. Our situation does not correspond to the Pfaffian, since we are dealing with the zero-set of a form considered as a section. A Pfaffian usually has constant rank, so its zero-set would be empty. For more information on Pfaffian systems, cf. \cite{Libermann87}.

The notion of \Def{def:generating} is equivalent to that of \Eq{eq:generating}. Indeed, if $j:M\to N$ represents the embedding $j(m)=(m,f(m))$, note that $\pi_1 \circ j=id_M$ and $\pi_2 \circ j=f$. The implication is that a generated map is exact.

\begin{pro}
If $\Lambda$ generates a map $f$ with respect to $(\alpha,\tilde \alpha)$, then $f$ is
exact volume-preserving with 
\begin{equation}
	f^*\tilde{\alpha}-\alpha =d \lambda \;.
\end{equation}
where $\lambda=j^*\Lambda$.
\end{pro}

Another implication of \Def{def:generating} is that the resulting map is invertible. Moreover, there is a simple relation between the generating function of a map $f$ and  its inverse.

\begin{pro}
If $f$ is generated by $\Lambda_f$ and $f$ is invertible then the inverse is generated by
\[
	\Lambda_{f^{-1}}=-\sigma^*\Lambda_f \;,
\]
where $\sigma:N\to N$ is the permutation $\sigma(m_1,m_2)=(m_2,m_1)$.
\end{pro}

Since by assumption $d\tilde\alpha = d\alpha = \Omega$, the difference $\tilde\alpha - \alpha$ is closed. If this difference is also exact, then $\tilde\alpha = \alpha + d\beta$. In this case, if $f$ is exact volume preserving with respect to the pair $(\alpha,\tilde\alpha)$, it is also exact with respect to $(\alpha,\alpha)$: 
$f^*\alpha-\alpha = f^*(\tilde\alpha -d\beta) -\alpha = d(\lambda-f^*\beta)$.
This same property holds for generating forms.

\begin{lem}[Legendre Transformations]\label{lem:change}
If $\Lambda $ is a generating form for $f$ with respect to $(\alpha,\tilde{\alpha})$ then $\Lambda+\pi_2^*\tilde{\beta}-\pi_1^*\beta$ is as well, with respect to the pair $(\alpha+d\beta,\tilde{\alpha}+d\tilde{\beta})$ for any $(n-2)$-forms $\beta$ and $\tilde{\beta}$.
\end{lem}
\proof
It is enough to notice from \Eq{eq:Gamma} that
\begin{equation}\label{eq:Legendre}
	\Gamma=
	\pi_2^*\left(\tilde{\alpha}+d\tilde{\beta}\right)-
	\pi_1^*\left(\alpha+d\beta\right)-d\left(\Lambda+\pi_2^*\tilde{\beta}-\pi_1^*\beta\right).
\end{equation}
\qed

This property is analogous to the Legendre transformations between various symplectic generating functions \cite{Arnold78, Goldstein02}. For example in $\Rset^n$, any \emph{even} permutation, 
\[
	p_{(i)}(x_1,x_2,\ldots,x_n)=(x_{i_1},x_{i_2},\ldots, x_{i_n}) \;,
\]
is an exact volume-preserving diffeomorphism. Thus, if $\alpha_{(i)} = p_{(i)}^*\alpha$, then there is an $(n-2)$-form $\beta_{(i)}$ such that $\alpha_{(i)}-\alpha = d\beta_{(i)}$. Consequently, if $\Lambda$ is a generates $f$ with respect to $(\alpha,\alpha)$, then \Eq{eq:Legendre} gives new generators with the permuted forms
\[
	\pi_2^*\alpha_{(i)} - \pi_1^*\alpha_{(j)}= d \Lambda_{(i),(j)}
\]
where
\begin{equation}\label{eq:permuted}
	\Lambda_{(i),(j)} = \Lambda + \pi_2^* \beta_{(i)} - \pi_1^* \beta_{(j)} \;.
\end{equation}
In this way, beginning with a basic form, say $\alpha = x_1 dx_2 \wedge \ldots \wedge dx_n$, and an associated generator, we can obtain generators for the $\frac12 n!$ evenly permuted forms 
$\alpha_{(i)} = x_{i_1} dx_{i_2} \wedge \ldots \wedge x_{i_n}$. Since each permutation can be done on each copy of $M$, there are $(\frac12 n!)^2$ possibilities.

\section{Thirty-Six Generating Forms on $\Rset^3$}\label{sec:thirtysix}

For $\Rset^3$, we will begin with the basic form $z dx \wedge dy$, and by even permutation
construct the two additional forms  $x dy \wedge dz$,  and $y dz \wedge dx$. Since any of the three can be used as well for $\tilde \alpha$, there are nine choices for the form $\Gamma$. For each such choice, we will see that there are four possible representations for $\Lambda$. Thus overall we will find thirty-six different generating forms. These are analogous to the four basic generating functions for area-preserving maps \cite{Arnold78,Goldstein02}.

To catalog the possibilities, begin by choosing $\tilde\alpha = \alpha = z dx \wedge dy$, and consider the generating equation
\begin{equation}\label{eq:generatorZero}
	\Gamma = Z dX \wedge dY - z dx \wedge dy - d\Lambda = 0
\end{equation}
on the graph of a smooth function $F = \{(m,f(m))\in\Rset^3\times\Rset^3 \;|\; m\in\Rset^3\}$.
A general one-form on $N$ will have terms for each of the six coordinates $(x,y,z,X,Y,Z)$; however, to be consistent with \Eq{eq:generatorZero}, $d\Lambda$ can have no terms involving $dz$ and $dZ$. This implies that any $z$ and $Z$ dependence of $\Lambda$ can be collected into terms that are total differentials: these give no contribution to the determination of $f$. Consequently, we set $\Lambda = Adx + Bdy + CdX + DdY$ where the functions
$A,B,C$, and $D$ depend only upon the four variables $(x,y,X,Y)$. Substitution into \Eq{eq:generatorZero} then gives six equations. The first two are dynamical in nature,
\begin{align*}
	z &= \partial_y A - \partial_x B \;, &
	Z &= \partial_X D - \partial_Y C \;,
\end{align*}
and the last four are the implicit consistency equations
\begin{align*}
	\partial_X A &= \partial_x C  \;, &
	\partial_Y A &= \partial_x D \;,\\
	\partial_X B &= \partial_y C \;, &
	\partial_Y B &= \partial_y D \;.
\end{align*}
There is a redundancy in these consistency equations that can be traced to the definition of $\Lambda$. Indeed, if we were to impose any one of these equations from the outset, we could rewrite $\Lambda$ as a form containing only two terms, up to a perfect differential. For example, the first consistency equation implies that $Adx + CdX = d\zeta - \partial_y\zeta dy - \partial_Y \zeta dY$ where---since $\partial_X A = \partial_x C $---we can set $\zeta = \int C dX = \int A dx$. Since $d\zeta$ does not enter into the generating equation \Eq{eq:generatorZero}, $A$ and $C$ can be effectively eliminated so that $\Lambda$ becomes $Bdy + DdY$. Two of consistency equations now reduce to $\partial_X B = \partial_x D  = 0$, which implies
$
	\Lambda = B(x,y,X) dy + D(y,X,Y)dY
$.
There remain three equations to implicitly determine the three components of the map $(X,Y,Z) = f(x,y,z)$:
\[
	z = - \partial_x B(x,y,X) \;, \quad
	\partial_Y B(x,y,X) = \partial_x D(y,X,Y) \;, \quad
	Z = \partial_X D(y,X,Y) \;.
\]
This map is well-defined only if these implicit equations can be inverted. The first equation can be solved for $X(x,y,z)$ only if $\partial_{xX}B \neq 0$, and the second can then be solved for $Y(x,y,z)$ only if $\partial_{xY} D \neq 0$. More specifically $f$ must satisfy two conditions: the curves $C = \{X(x,y,z) \;|\; z \in \Rset\}$, and $\tilde C= \{y(X,Y,Z) \;|\; Z \in \Rset\}$ must be bijections onto $\Rset$ for each fixed $(x,y)$ and $(X,Y)$, respectively. This will occur, for example, if the derivatives $\partial{X}/\partial{z}$ and $\partial{y}/\partial{Z}$ are uniformly positive and bounded:
\[
	0 < \ell_1 \le \frac{\partial X}{\partial z}\;, \; 
	    \frac{\partial y}{\partial Z} \le \ell_2 < \infty \;.
\]

A similar reduction of $\Lambda$ to two terms can be performed by imposing each of the remaining three consistency equations, giving four basic generating forms as shown in \Tbl{tbl:fourBasic}. These four are geometrically distinct in that they have distinct twist conditions.

\begin{table}[ht]
\centering
\begin{tabular}{c|c|c}
$\Lambda_{0,0}$  & $Adx$ & $Bdy$ \\
\hline
       & $A(x,y,X) \;,\quad C(x,X,Y)$ & $B(x,y,X) \;, \quad C(y,X,Y)$ \\
	   & $z=\partial_yA$ 		 & $z=-\partial_xB$\\
$C dX$ & $\partial_X  A = \partial_x C$ 	  & $\partial_X B = \partial_y C$\\
	   & $Z = -\partial_Y C$ 	  & $Z = -\partial_Y C$ \\
	   & $\frac{\partial X}{\partial z} \neq 0\;, \quad \frac{\partial x}{\partial Z} \neq 0$	  
	   & $\frac{\partial X}{\partial z} \neq 0\;, \quad \frac{\partial y}{\partial Z} \neq 0$ \\
\hline
       & $A(x,y,Y) \;, \quad D(x,X,Y)$   & $B(x,y,Y)\;, \quad D(y,X,Y)$\\
	   & $z=\partial_yA$ 		 & $z=-\partial_xB$\\
$D dY$ & $\partial_Y  A = \partial_x D$ 	  & $\partial_Y B = \partial_y D$\\
	   & $Z = \partial_X D$ 	  & $Z = \partial_X D$ \\
	   & $\frac{\partial Y}{\partial z} \neq 0 \;, \quad \frac{\partial x}{\partial Z} \neq 0$	  
	   & $\frac{\partial Y}{\partial z} \neq 0\;, \quad \frac{\partial y}{\partial Z} \neq 0$ \\
\end{tabular}
\caption{\footnotesize Four basic generating forms with respect to $\tilde \alpha = \alpha = zdx\wedge dy$. Shown are the independent variables for each function, the three implicit mapping equations, and the two twist conditions.
\label{tbl:fourBasic}}
\end{table}

Additional generating forms can be obtained from \Tbl{tbl:fourBasic} using the Legendre transformation \Eq{eq:permuted} to change the forms $\pi_2^*\alpha = Z dX \wedge dY$ and $\pi_1^*\alpha = z dx \wedge dy$ into the eight remaining permutations. Specifically let $p_{(231)}(x,y,z) = (y,z,x)$ and $p_{(312)}(x,y,z) = (z,x,y)$ denote the even permutations. Then 
\begin{align*}
	p_{(231)}^*\alpha -\alpha &=  x dy \wedge dz - z dx \wedge dy = d(-xz dy) \;, \\
	p_{(312)}^*\alpha -\alpha &=  y dz \wedge dx - z dx \wedge dy = d(yz dx) \;.
\end{align*}
Thus, the generator for 
\[
	\pi_2^* \alpha - \pi_i^* p_{(231)}^*\alpha = Z dX\wedge dY - x dy\wedge dz \;,
\]
becomes
\[
	\Lambda_{(231),0} = \Lambda_{0,0} + xz dy \;.
\]
For example, to reproduce the results of \Sec{sec:Carroll} we select the $dy$ and $dY$ components for $\Lambda$, so we begin with the $B$-$D$ form
for $\Lambda_{0,0}$ to obtain
\[
	\Lambda_{(231),0} = (B(x,y,Y)+xz)dy + D(y,X,Y)dY = \hat A(y,z,Y) dy + D(y,X,Y) dY
\]
As indicated, the mapping equation $B_x = -z$, becomes a new consistency condition: $\partial_x \hat A = 0$.  The consistency condition $\partial_z B = 0$ becomes a new mapping equation
\[
	\partial_z \hat A = \partial_z B + x = x \;.
\]
The remaining two equations are unchanged, reproducing the system \Eq{eq:carrollMap}.
Alternatively, the permutation can also directly be applied to the labels $(x,y,z)$ in \Tbl{tbl:fourBasic} to transform the entire table into that for $\Lambda_{(231),0}$, see \Tbl{tbl:fourPermuted}. Note that the twist conditions for the generated maps are geometrically distinct.

Similar tables are easily constructed for the remaining permutations to give a total of thirty-six different generating forms.

\begin{table}[ht]
\centering
\begin{tabular}{c|c|c}
$\Lambda_{(231),0}$  & $Ady$ & $Bdz$ \\
\hline
       & $A(y,z,X) \;,\quad C(y,X,Y)$ & $B(y,z,X) \;, \quad C(z,X,Y)$ \\
	   & $x=\partial_zA$ 		 & $x=-\partial_yB$\\
$C dX$ & $\partial_X  A = \partial_y C$ 	  & $\partial_X B = \partial_z C$\\
	   & $Z = -\partial_Y C$ 	  & $Z = -\partial_Y C$ \\
	   & $\frac{\partial X}{\partial x} \neq 0\;, \quad \frac{\partial y}{\partial Z} \neq 0$	  
	   & $\frac{\partial X}{\partial x} \neq 0\;, \quad \frac{\partial z}{\partial Z} \neq 0$ \\
\hline
       & $A(y,z,Y) \;, \quad D(y,X,Y)$   & $B(y,z,Y)\;, \quad D(z,X,Y)$\\
	   & $x=\partial_zA$ 		 & $x =-\partial_yB$\\
$D dY$ & $\partial_Y  A = \partial_y D$ 	  & $\partial_Y B = \partial_z D$\\
	   & $Z = \partial_X D$ 	  & $Z = \partial_X D$ \\
	   & $\frac{\partial Y}{\partial x} \neq 0 \;, \quad \frac{\partial y}{\partial Z} \neq 0$	  
	   & $\frac{\partial Y}{\partial x} \neq 0\;, \quad \frac{\partial z}{\partial Z} \neq 0$ \\
\end{tabular}
\caption{\footnotesize Four basic generating forms with respect to $\tilde \alpha = zdx \wedge dy$ and  $\alpha = xdy \wedge dz$. \label{tbl:fourPermuted}}
\end{table}

As an example of the forms shown in \Tbl{tbl:fourPermuted}, 
consider the $B$-$C$ type generating form
\[
	\Lambda = (-yX + g(y,z) - h(X,z))dz + (-zY - k(X,Y))dX \;.
\]
The generated map is
\begin{equation}\label{eq:abcMap} \begin{split}
	X &= x + g_y(y,z) \;,\\
	Y &= y + h_X(X,z) \;,\\
	Z &= z + k_Y(X,Y) \;.
\end{split}\end{equation}
The twist conditions are trivially satisfied: $\partial_x X(x,y,z) = 1$ and 
$\partial_Z z(X,Y,Z) = 1$. The much-studied $ABC$-map has this form \cite{Feingold88, Piro88}. The map \Eq{eq:abcMap} is the composition of three, exact volume-preserving shears, e.g., maps of the form $(X,Y,Z) = (x+F(y,z),y,z)$. It is also a first-order volume-preserving integrator of the incompressible flow with vector field $(g_y(y,z),h_x(x,z),k_y(x,y))$ \cite{McLachlan01}.

\section{Some Generating forms on $\Rset^n$}

In this section we construct a generating form for $M = \Rset^n$, choosing---for simplicity,
\begin{align*}
	\pi_1^*\alpha         &= (-1)^{n-1}x_ndx_1\wedge\cdots\wedge dx_{n-1}, \\
	\pi_2^*\tilde{\alpha} &= X_1 dX_2\wedge\cdots\wedge dX_{n} \;.
\end{align*}	
Here we use the coordinates $(x_1,\ldots,x_n,X_1,\ldots, X_n) \in N = M \times M$.
This choice will reproduce formulas that, as far as we know, first appeared in \cite{Carroll04}.

The form $\Lambda$ will depend upon the $n-2$ variables
$(x_1,x_2,\ldots x_{n-1},X_2,X_3,\ldots,X_n)$. To develop the notation for this form,
define the projections $h_k:N\to M$  by
\begin{equation}
	h_k(x_1,\ldots,x_n,X_1,\ldots, X_n)=(x_1,\ldots,x_k,X_{k+1},\ldots,X_n),
\end{equation}
for each $k=1,\ldots,n-1$. Similarly for each $k$, define the $(n-2)$-form on $M$
\begin{align*}
	\rho_k &\equiv \Phi^k\,dx_1\wedge\cdots\wedge dx_{k-1}\wedge 
				dx_{k+2}\wedge\cdots\wedge dx_{n}
\end{align*}
where $\Phi^k \in C^2(M,\Rset)$. 
Notice that $h_k^*\rho_k$ is an $(n-2)$-form defined on $N$.

\begin{teo}
Let $\Phi^1,\ldots,\Phi^n$ be smooth functions on $M$. Assume that there exist two constants
$\ell_1,\ell_2>0$ such that, for all $k=1,\ldots,n-1$ and all $m\in M$, one has
\[
	0<\ell_1\leq\left|\partial_{k,k+1}\Phi^k(m)\right|\leq\ell_2 \;.
\]
Then, the $n-2$-form $\displaystyle \Lambda=\sum_{k=1}^{n-1}h_k^*\rho_k$ is a generating form
and the
generated map $(X_1,\ldots,X_n)=f(x_1,\ldots,x_n)$ is implicitly given by the $n$ equations
  \begin{equation}\label{eq:xxx}\begin{split}
		X_1 &= \partial_2\Phi^1(x_1,X_2,\ldots,X_n) \;, \\
	    \partial_k\Phi^k(x_1,\ldots,x_k,X_{k+1},\ldots,X_n) &= 
	    		\partial_{k+2}\Phi^{k+1}(x_1,\ldots,x_{k+1},X_{k+2},\ldots,X_n)  \;,\\
		\partial_{n-1}\Phi^{n-1}(x_1,\ldots,x_{n-1},X_n) &= x_n  \;.\\
  \end{split}\end{equation}
  for $k=1,\ldots,n-2$.
\end{teo} 
\proof
This is a straightforward computation. The differentials of the basic forms are
\[\begin{split}
	d\rho_k
	= &(-1)^{k-1}\left(\partial_k\Phi^k \right)dx_1\wedge\cdots\wedge dx_{k}
			\wedge dx_{k+2}\wedge\cdots\wedge dx_{n}\\
	&+ (-1)^{k-1}\left(\partial_{k+1}\Phi^k\right)dx_1\wedge\cdots\wedge dx_{k-1}
			\wedge dx_{k+1}\wedge\cdots\wedge dx_{n} \;.
\end{split}\]
This implies that, as a form on $N=M\times M$, $\Lambda$ satisfies: 
\begin{align*}
	d\Lambda=\sum_{k=1}^{n-1} (-1)^{k-1} &\left[
		\left(\partial_k\Phi^k \circ h_k\right)
		 dx_1\wedge\cdots\wedge dx_{k}\wedge dX_{k+2}\wedge\cdots\wedge dX_{n} \right. \\
	&+ \left. \left(\partial_{k+1}\Phi^k\circ h_k\right)dx_1\wedge\cdots\wedge dx_{k-1}
		\wedge dX_{k+1}\wedge\cdots\wedge dX_{n} \right] \;.
\end{align*}
Rearranging the terms in the sum we find that
\begin{align*}
	d\Lambda =& 
		\left(\partial_{2}\Phi^{1}\circ h_{1}\right)dX_{2}\wedge\cdots\wedge dX_{n}
	+ (-1)^{n}\left(\partial_{n-1}\Phi^{n-1}\circ h_{n-1}\right)dx_1\wedge\cdots\wedge dx_{n-1} \\
   +&\sum_{k=1}^{n-2}(-1)^{k}\left(\partial_{k+2}\Phi^{k+1}\circ h_{k+1}-
		\partial_k\Phi^k \circ h_k\right)
	 dx_1\wedge\cdots\wedge dx_{k}\wedge dX_{k+2}\wedge\cdots\wedge dX_{n} \;.
\end{align*}
Therefore, in order to satisfy 
\[
 \left(\pi_2^*\tilde{\alpha}-\pi_1^*\alpha-d\Lambda\right)(x_1,\ldots,x_n,X_1,\ldots, X_n)=0 \;,
\]
one needs to have,   for $k=1,\ldots,n-2$,
\begin{equation}\label{eq:yyy}\begin{split}
	X_1-\partial_{2}\Phi^{1}\circ h_{1}&= 0 \;, \\ 
	\partial_{k+2}\Phi^{k+1}\circ h_{k+1}- \partial_k\Phi^k \circ h_k &= 0 \;,\\ 
	x_n -\partial_{n-1}\Phi^{n-1}\circ h_{n-1} &= 0 \;.
\end{split}\end{equation}
Equations (\ref{eq:xxx}) and (\ref{eq:yyy}) are the same. The conditions on the functions $\Phi^k$ imply that we can solve for $(X_1,\ldots,X_n)$ in terms of $(x_1,\ldots,x_n)$ and vice versa.
\qed

For example, on $\Rset^3$ with coordinates $(x,y,z)$, this reduces to the generating form
\[
	\Lambda_{0,(231)} = \Phi^1(x,Y,Z)dZ + \Phi^2(x,y,Z)dx \;,
\]
which is a permuted version of the $A$-$D$ form in \Tbl{tbl:fourBasic}.

\section{Generators on Other Manifolds}\label{sec:manifolds}

When $M$ has nontrivial homology, the volume form $\Omega$ on $M$ need not be exact, in which case $\alpha$ cannot even be defined. Even if an $(n-1)$-form $\alpha$ can be defined, there might not be sufficient freedom to select these forms to obtain a well-defined generator $\Lambda$ on $M\times M$.  However, in this case, we may still be able to find a generator on the universal cover of $M$ since it is simply connected.

Suppose that $p: C \to M$ is a cover of $M$ and that the volume form $p^*\Omega$ on $C$ is exact. In this case, even if the original $\Omega$ is not exact, it is possible to find generating forms on the universal cover.

We now construct $\Lambda$ on the product space for the universal cover $N = C \times C$. If $\Lambda$ is a generating form on $N$, then it generates a map $g : C \to C$. This map will be the lift of a map $f: M \to M$ if
\[
	p \circ g = f \circ p \;.
\]
We will show that for this to occur it is sufficient that $\Gamma$ be invariant under an extension of the group of deck transformations of $p$ to the space $C\times C$.

Recall that the group of deck transformations of a cover $p$ is $T = \{ t: C \to C \;|\; p \circ t = p\}$. Obviously $p\times p:C\times C\to M\times M$ is a cover of $M\times M$ and the group of deck transformations on $C\times C$ is
\[
U=\{(t_1,t_2):C\times C\to C\times C \;|\; t_1,t_2\in T\} \;;
\]
indeed, if $(t_1,t_2)\in U$ then $(p,p) \circ (t_1,t_2) =(p,p)$.
We will say that a subgroup $\Delta$ of $U$ is of \emph{diagonal type} if it is of the form
\[
	\Delta = \{(t,\psi(t))\in U \;|\; t\in T\}
\]
for some fixed group automorphism $\psi:T\to T$. For instance, if we take $\psi=id_T$, then $\Delta$ is the diagonal. In general, $\Delta$ is isomorphic to the original group $T$. Invariance of $\Lambda$ under $\Delta$ implies that its generated map is a lift. 

\begin{lem}\label{lem:bonito}
Suppose that $C$ is a cover of $M$, $\alpha$ and $\tilde{\alpha}$ are $n-1$ forms in $C$ such that $p^*\Omega=d\alpha=d\tilde{\alpha}$, and $\Lambda$ is a generator with respect to $(\alpha, \tilde\alpha)$ of an exact volume-preserving map $g$ on $C$.
Then, if $\Gamma = \pi_2^*\tilde\alpha-\pi_1^*\alpha - d\Lambda$ is invariant under a subgroup of the deck transformations of $C\times C$ of diagonal type, $g$ is the lift of a volume-preserving map $f$ on $M$. 	In addition, if $\Omega$ is exact, then $f$ is exact volume-preserving.
\end{lem} 

\proof

By assumption $\Gamma$ vanishes at $(c, g(c)) \in N$ for each $c \in C$. Since $\Gamma$ is invariant under $\Delta$, $(t,\psi(t))^*\Gamma = \Gamma$. Therefore $\Gamma$ also vanishes at $(t(c),\psi(t)(g(c))$. Equivalently, 
\[
  g \circ t  = \psi(t) \circ g \;.
\]
Consequently, the two points $t(c)$ and $c$, which project to the same point $m = p(c)$, have the same projected image $p(g(t(c)) = p(g(c))$. But this implies that $f(m)$ is uniquely defined and satisfies $p\circ g= f\circ p$.

Now $p^*\Omega$ is a volume form on $C$ and $g$ preserves $p^*\Omega$. Hence
\begin{equation}
	p^*f^*\Omega = (f\circ p)^*\Omega = (p\circ g)^*\Omega=p^*\Omega.
\end{equation}
Since $p$ is a local diffeomorphism, we conclude that $f^*\Omega=\Omega$. When the original volume form is exact, the same argument can be used to show that $f$ is also exact.

\qed

Note that even when $\Gamma$ is invariant under $\Delta$, the form $\Lambda$ need be, and thus may not have a well-defined projection on $M\times M$. Nevertheless, the projected map is well-defined whenever $\Gamma$ is invariant under a suitable subgroup $\Delta$.

\begin{example}
A simple example corresponds to the two-dimensional, generalized standard map
\begin{equation}\label{eq:stdMap}
	f(x,y) = (x + y - V'(x), y - V'(x)) \;.
\end{equation}
where $V(x+1) = V(x)$ is the potential. We can think of this map as being defined on the cylinder $M = \Tset \times \Rset$. In this case, 
the universal cover is $C = \Rset^2$. The group $T$ of deck transformations is generated by a single transformation: $T=\langle \phi_1 \rangle$, where $\phi_1(x,y)=(x+1,y)$. Using the volume form $\Omega = dy \wedge dx$, we may select $\alpha = \tilde \alpha = y dx$, and obtain the generating form
\begin{equation}\label{eq:StdMapGen}
	\Lambda = \frac12 (X -x)^2 - V(x) \;,
\end{equation}
which is a zero-form on $C\times C$, but not on $M \times M$. On the manifold $C\times C$, we will use the
the diagonal extension $\Delta$ of $T$ that consists of the transformations $u(x,y,X,Y) = (x+k,y,X+k,Y)$ for integer $k$. In other words, we simple use $\psi=id_T$ in the argument above. It enough to check that the form $\Gamma= YdX - ydx-d\Lambda$ is invariant under
$(\phi_1,\phi_1)$.  Thus the generated map projects to an exact volume-preserving  map on $\Tset \times \Rset$.

We can also think of \Eq{eq:stdMap} as acting on $M = \Tset^2$. In this case $dy \wedge dx$ is closed, but not exact: $\alpha = y dx$ is not a form on $M$.
The universal cover is still $\Rset^2$; however, 
the deck transformation group is now $T = \{t(x,y) = (x+m,y+n) \;|\; m,n \in \Zset\}$ which is generated by two transformations: $T=\langle \phi_1,\phi_2 \rangle$, where
\begin{align*}
	\phi_1(x,y)&=(x+1,y) \;,\\
	\phi_2(x,y)&=(x,y+1) \;.
\end{align*}
The zero-form \Eq{eq:StdMapGen} is still a generator on the cover; however, the one form
\[
	\Gamma = YdX - ydx -d\Lambda 
			= (Y-X+x)dX - (y-X+x + V'(x))dx
\]
is not invariant under the  trivial diagonal extension of $T$. Instead, define a different subgroup $\Delta$ by choosing the automorphism $\psi$ so that $\psi(\phi_1)=\phi_1$ and $\psi(\phi_2)=\phi_1\circ\phi_2=\phi_2\circ\phi_1\equiv \phi_3$.
Now, the subgroup $\Delta$ is generated by the pairs
\begin{align*}
	u_1(x,y,X,Y) &=(\phi_1,\phi_1)(x,y,X,Y) = (x+1,y,X+1,Y) \;, \\
	u_2(x,y,X,Y) &=(\phi_2,\phi_3)(x,y,X,Y) = (x,y+1,X+1,Y+1) \;.
\end{align*}
The form $\Gamma$ is invariant under these deck transformations: $u_1^*\Gamma=\Gamma$ and $u_2^*\Gamma=\Gamma$. Thus every map $g:\Rset^2\to\Rset^2$ generated in this way has the symmetries:
\begin{align*}
	g\circ \phi_1&=\phi_1\circ g \;,\\
	g\circ \phi_2&=\phi_3\circ g \;.
\end{align*}
Therefore $g$ is the lift of a well-defined map of $\Tset^2$.

\end{example}

\section{One-Action Maps}\label{sec:oneAction}

An action-angle map $f$ acts on the manifold $M = \Tset^d \times \Rset^k$, having $d$ angle variables, $\theta \in \Tset^d$, and $k$ action variables, $z \in \Rset^k$. As an example, consider the generalization of \Eq{eq:stdMap}
\begin{equation}\label{eq:oneActionMap}
	f(\theta,z) = \left(\theta + \rho(z + F(\theta)), z + F(\theta) \right)
\end{equation}
for a  ``rotation vector'' $\rho: \Rset^k \to \Tset^d$ and ``force'' $F: \Tset^d \to \Rset^k$. Examples with two angles and one action have been much studied, cf. \cite{Feingold88, Cartwright96}. Here we will consider this case, setting $d=2$ and $k=1$ and 
\begin{equation}\label{eq:oneActionForm}
	\Omega = dz \wedge d\theta_1 \wedge d\theta_2 \;, \quad 
	\alpha =  z d\theta_1 \wedge d\theta_2 \;.
\end{equation}

The map \Eq{eq:oneActionMap} is volume preserving. It is easiest to see this by noting that $f = f_1 \circ f_2$  for the volume-preserving shears $f_1(\theta,z) = (\theta+\rho(z),z)$ and $f_2(\theta,z) = (\theta,z+F(\theta))$. Moreover, the map $f_1$ is always exact, it satisfies \Eq{eq:exact} with $\lambda_1 =  i_W \,d\theta_1 \wedge d\theta_2 = W_1 d\theta_2 -W_2 d\theta_1$ with $W_i = z\rho_i - \int \rho_i dz$. By contrast, $f_2$ is exact only when
\[
	\int_{\Tset^d} F(\theta) \,d\theta_1 \wedge d\theta_2= 0 \;.
\]
This is true if there a vector field $G :\Tset^2 \to \Rset^2$ such that $\nabla \cdot G = F$. 
In this case the $\lambda_2 =  i_G d\theta_1 \wedge d\theta_2$. Finally, the form $\lambda$ for $f$ is defined using \Eq{eq:composition}.

A \emph{rotational} torus is a two-dimensional torus homotopic to the zero section $\{ (0,\theta) \;|\; \theta \in \Tset^2 \}$. The \emph{net flux} crossing a rotational torus $\cT$ is the difference between the volume ``below'' $f(\cT)$ and that below $\cT$:
\begin{equation}\label{eq:netFlux}
	\cF(\cT) = \int_{\cT} f^*\alpha-\alpha   \;.
\end{equation}
When $f$ is exact, then $\cF(\cT) =0$. A consequence is that $f(\cT) \cap \cT \neq \emptyset$. In fact, this intersection property has been used in generalizing KAM theory for exact volume-preserving maps \cite{Xia92}.

The natural integrable case of \Eq{eq:oneActionMap} is 
\begin{equation}\label{eq:integrable}
	f(\theta, z) = (\theta + \rho(z), z) \;,
\end{equation}
For this map, the phase space is foliated by invariant two-tori. Interestingly under some conditions on $\rho$, a version of KAM-theory can be applied to this system to imply that a Cantor-set of these tori are preserved when $f$ is smoothly perturbed, but remains exact volume-preserving \cite{Sun84,Xia92}.

Indeed, exactness is a necessary requirement for the existence of rotational invariant tori. Suppose that $\cT$ and $\hat\cT$ are rotational tori and $f$ is volume-preserving, then the volume contained between them, $\Delta V = \int_{\hat\cT} \alpha - \int_{\cT} \alpha$, is invariant. This implies that the flux $\cF$ is independent of the choice of torus. Consequently if $f$ has an invariant torus then its net flux must vanish, $\cF = 0$. Since $\cF(\cT)= 0$ for any rotational torus, $f^*\alpha-\alpha  $ must be exact. Therefore a necessary condition for the existence of rotational invariant tori is that $f$ be exact volume-preserving.


When $M=\Tset^2\times\Rset$, there is only one choice for $\alpha$ that will make the standard volume form exact, \Eq{eq:oneActionForm}. Thus, we will consider a generating equation of the form
\[
\begin{split}
	\Gamma&=\pi_2^*\alpha -\pi_1^*\alpha-d\Lambda\\
	   &= Z d\Theta_1 \wedge d\Theta_2 - z d\theta_1 \wedge d\theta_2 
	  -d\Lambda \;.
\end{split}
\]
Taking $\Tset^2 = \Rset^2 / \Zset^2$, we can define the generating form $\Lambda$ on the universal cover, $C=\Rset^3$, of $M$. Using $(\theta, z, \Theta, Z)$ as coordinates on $\Rset^3 \times \Rset^3$, a suitable generating form is
\begin{equation}\label{eq:oneActionGen}
	\Lambda=\left[(\Theta_1-\theta_1)Z  - \Phi(\theta_1,\Theta_2,Z) \right]d\Theta_2
	     - \left[(\Theta_2-\theta_2)z - \Psi(\theta_1,z,\Theta_2) \right]d\theta_1.
\end{equation}

The appropriate diagonal extension of the group of deck transformations of $C$ consists
of the transformations
\[	
	u_k(\theta,z, \Theta,Z) = (\theta + k,z,\Theta +k,Z)\;, \quad k \in \Zset^2 \;.
\]
In order for $\Lambda$ to generate the lift of a map on $M$, it must be invariant: $u_k^*\Lambda = \Lambda$. This requirement is easily seen to be satisfied by \Eq{eq:oneActionGen} when $\Phi$ and $\Psi$ are periodic in each of their angular arguments.

The differential of the generating form \Eq{eq:oneActionGen} is
\begin{align*}
	d\Lambda =& \left(z-Z - \partial_{\theta_1}\Phi
			-\partial_{\Theta_2}\Psi\right)d\theta_1\wedge d\Theta_2\\
			  &+ \left(\Theta_1-\theta_1-\partial_Z\Phi \right)dZ\wedge d\Theta_2
			  + \left(\theta_2-\Theta_2 +\partial_z\Psi\right)dz\wedge d\theta_1 \\
			  &+ Z d\Theta_1 \wedge d\Theta_2 - z d\theta_1\wedge d\theta_2 \;.
\end{align*}
Note that in this case $d\Lambda$ includes the terms in $\pi_2^*\alpha-\pi_1^*\alpha$. The remaining three terms in $d\Lambda$ must vanish, and this determines the generated map
\begin{align*}
	\Theta_1 &= \theta_1 + \partial_Z\Phi(\theta_1,\Theta_2,Z)  \;,\\
	\Theta_2 &= \theta_2 + \partial_z\Psi(\theta_1,z,\Theta_2)  \;,\\
	Z        &= z - \partial_{\theta_1}\Phi(\theta_1,\Theta_2,Z)
	       -\partial_{\Theta_2}\Psi(\theta_1,z,\Theta_2) \;.
\end{align*}
This map is one-to-one when $\partial_{z\Theta_2} \Psi \neq 1$ and $\partial_{\theta_1 Z}\Phi \neq -1$. This map will have the form of a perturbation of the integrable map \Eq{eq:integrable} if we set
\begin{align*}
	\Phi &= H_1(Z) + \epsilon F(\theta_1, \Theta_2) \;, \\
	\Psi &= H_2(z) + \epsilon G(\theta_1, \Theta_2) \;,
\end{align*}
which gives the semi-explicit map
\begin{align*}
	\Theta_1 &= \theta_1 + \partial_Z H_1(Z)  \;,\\
	\Theta_2 &= \theta_2 + \partial_z H_2(z)  \;,\\
	Z        &= z - \epsilon[\partial_{\theta_1}F(\theta_1,\Theta_2)
	              +\partial_{\Theta_2}G(\theta_1,\Theta_2)] \;.
\end{align*}

\section{Conclusions}\label{sec:conclusion}

We have shown that exact volume-preserving maps on an $n$-dimensional manifold can have implicit generating $(n-2)$-forms in the same way that symplectic maps can have implicit generating functions (zero forms). In both cases, the generated maps must satisfy certain necessary geometrical conditions that we have called \emph{twist conditions}. For the $n$-dimensional case, there are $n-1$ such conditions. It would be interesting to characterize the exact volume-preserving diffeomorphisms that can be generated in this way in terms of a suitable set of twist conditions.

One of the reasons for defining generating functions is to obtain variational principles. These are used to great effect, for example, in Aubry-Mather theory for area-preserving twist maps \cite{Meiss92}. Variational principles for exact incompressible vector fields  have been studied by Gaeta and Morano \cite{Gaeta03} (where they were called ``globally Liouville vector fields''). It would be interesting to extend their analysis to the map case.

Another possible use for generating forms is as integration algorithms for incompressible flows. Implicit generators are commonly used in symplectic integration algorithms \cite{McLachlan01}.

Generating functions for symplectic maps are also used to compute the symplectic area of lobes in the theory of transport. We will generalized this result to the exact volume-preserving case in a forthcoming paper \cite{Lomeli08b}.

\bibliographystyle{alpha}
\bibliography{tev1}

\end{document}